\documentclass[conference]{IEEEtran}

\input{psfig}
\newcommand{\Se}[1]{\section{#1}\label{#1}}
\newcommand{\Sse}[1]{\subsection{#1}\label{#1}}
\newcommand{\Ssse}[1]{\subsubsection{#1}\label{#1}}

\newcommand{\Bi}{\begin{itemize}}
\newcommand{\Ei}{\end{itemize}}
\newcommand{\Bn}{\begin{enumerate}}
\newcommand{\En}{\end{enumerate}}
\newcommand{\Bd}{\begin{description}}
\newcommand{\Ed}{\end{description}}
\newcommand{\Bq}[1]{\begin{equation}\label{#1}}
\newcommand{\Eq}{\end{equation}}
\newcommand{\Bqn}[1]{\begin{eqnarray}\label{#1}}
\newcommand{\Eqn}{\end{eqnarray}}
\newcommand{\BTab} { \vspace{0.3cm} \begin{table}[phtb] \centering }
\newcommand{\ETab}[2] {\caption{#2} \label{#1:tab} \end{table} \vspace{0.3cm} }
\newcommand{\BTAB} { \vspace{0.3cm} \begin{table*}[phtb] \centering }
\newcommand{\ETAB}[2] {\caption{#2} \label{#1:tab} \end{table*} \vspace{0.3cm} }

\newcommand{\junk}[1] {}
\newif\ifsizedfigures \sizedfiguresfalse
\newif\ifincludefigures \includefiguresfalse
\newif\iffigurenamesinmargin \figurenamesinmarginfalse
\newif\ifpdf\ifx\pdfoutput\undefined\pdffalse\else\pdfoutput=1\pdftrue\fi

\usepackage{times}
\newcommand{\mycomment}[1]{}

\newcommand{\figw}[3]
{
\begin{figure}
\centerline{\psfig{figure=fig/#1.eps,width=#2}}
\caption{\label{#1:fig} \small #3}
\end{figure}
}

\begin{document}
\title{\bf Programmable Ethernet Switches and Their Applications 
}
\author{
Srikant Sharma\ \ \ Tzi-cker Chiueh \\
{\small \{srikant,chiueh\}@cs.sunysb.edu}    \\
Department of Computer Science \\
Stony Brook University, Stony Brook, NY-11794\\ 
}

\date{}
\maketitle

%
%
%
%

\begin{abstract}
\mycomment{
Despite its popularity, the service providers of metro Ethernet itself are
cold to Ethernet deployment in the network core. The metro Ethernet services
are provided by tunneling Ethernet frames over traditional circuit switched technologies
like SONET, RPR, ATM, or switching technologies like MPLS, etc. Similarly,
the popularity Ethernet in Cluster and Storage networks is driven by
its interconnection simplicity and cost effectiveness. The deployments are
usually monolithic or limited to  a few switches with large port densities, 
exploiting the performance of the switch crossbars rather than the Ethernet technology itself.
The primary reason for this trend is the inability to explicitly choose
switching paths in Ethernet precluding traffic engineering to maximize the performance on a broader scale.
In this paper, we propose a programmable switching paradigm for Ethernet networks, which enables
MPLS like switching path selection with fault-tolerant capabilities paving way for traffic engineering
and QoS in Ethernet and facilitates efficient reliable multicast beneficial for cluster and storage networks.
Programmable switching promises multi-fold throughput gains with sub-second level recovery.
This mechanism is realized by relying on simple programmable features of commodity Ethernet switches
such as VLAN configurability, status monitoring, rate limiting, and  IGMP snooping. Since it does
not require any changes to commercially available Ethernet switches, it is readily deployable in today's
networks.
}
Modern Ethernet switches support many advanced features beyond route
learning and packet forwarding such as VLAN tagging, IGMP snooping,
rate limiting, and status monitoring, which can be controlled through
a programmatic interface. Traditionally, these features are mostly used
to statically configure a network. This paper proposes to apply them as
dynamic control mechanisms to maximize physical network link resources,
to minimize failure recovery time, to enforce QoS requirements, and to
support link-layer multicast without broadcasting. With these advanced
programmable control mechanisms, standard Ethernet switches can be used as
effective building blocks for metropolitan-area Ethernet networks (MEN),
storage-area networks (SAN), and computation cluster interconnects. We
demonstrate the usefulness of this new level of control over Ethernet
switches with a MEN architecture that features multi-fold throughput
gains and sub-second failure recovery time.

\end{abstract}

\Se{Introduction}
Ethernet technology has come a long way from its initial shared-media 10
Mbps capability to today's switched-media form providing throughput up to
several Gbps. Its simplicity, cost effectiveness,  and the
economies of scale have enabled it to make forays into practically
various scales of networks, including metropolitan area networks.
Recent technological
advances, such as Ethernet in the First Mile~\cite{8023ah}, which
enables subscribers to connect to an Ethernet-based core network over a wide
variety of media ranging from voice grade copper to multi-mode fiber,
further reinforce the case of metropolitan Ethernet network (MEN) architecture.
Bandwidth availability of up to 10~Gbps and micro-second
level message latencies also make Ethernet a low-cost alternative to
widely used cluster interconnects (CI) such as Myrinet~\cite{myrinet},
Quadrics~\cite{quadrics}, and Infiniband~\cite{infiniband}. Finally, with the increasing
momentum of IP storage, 
Ethernet-based storage area networks (SAN) are becoming  formidable challengers
against fiber channel-based SAN because of lower cost and additional flexibility. 
These new applications
of Ethernet elevate it from a LAN technology to a ubiquitous networking
technology.

Many Metro Ethernet services today are actually built on  
circuit switched technologies such as SONET, ATM, RPR, or
established packet switching technologies like MPLS etc.
These services are actually provided by means of tunnels set up 
over these physical carrier technologies. Deployment of standard  
Ethernet switches in the core is still very rare, because there are
several architectural deficiencies with switching in Ethernet.
First, Ethernet networks
use a spanning tree protocol (IEEE 802.1d)~\cite{8021d} to
establish a path between any pair of nodes. 
It is well known that
the spanning tree approach fails to exploit all the physical network resource.
In addition, failure of switches and links requires rebuilding the
of spanning tree, which is a lengthy process.  IEEE 802.1w~\cite{8021w},
the rapid spanning tree configuration protocol (RSTP),  mitigates this
problem by providing mechanisms to detect failures and quickly reconfigure
the spanning tree. However, the recovery period can still range from an
optimistic 10 milliseconds to more realistic multiple seconds {\em after} failure detection,
which is still not adequate for many applications.  
Second, Ethernet does not support any
mechanism akin to MPLS~\cite{mpls}, which allows the user to route packets/flows 
along a particular path. As a result, it is impossible to  
apply any traffic engineering technique~\cite{te-qos} to balance traffic load
across the network. Traffic engineering may not be useful in small
local area networks, but is very important in the context of MAN,
SAN, and CI. In particular, the ability to route traffic on a given route
can also greatly help enforcing QoS by leveraging the traffic 
prioritization scheme (IEEE 802.1p)~\cite{8021p}, which  
prioritizes certain classes of
traffic over others.



The key insight of this work is that modern Ethernet switches incorporate 
advanced network control mechanisms that are programmatically configurable 
and could be used to improve the aggregate throughput, the availability 
and the QoS.
Virtual LAN (VLAN)~\cite{8021q} technology 
provides a mechanism to a physical subnet into 
multiple broadcast domains to improve the security
and performance of LANs. Multiple spanning tree (MST) protocol~\cite{8021s}
makes it possible
to configure multiple spanning tree instances on 
a network, each associated with a distinct VLAN, to isolate traffic from one
another.  Many commercially available Ethernet switches also support IGMP
snooping~\cite{igmpsnoop} for making intelligent multicast forwarding
decision by examining the Layer-3 IP headers and  use the network
resources more efficiently. Finally, most Ethernet switches can  
limit the rate of incoming or outgoing packets over their physical interfaces.

Most of these advanced structuring mechanisms in modern Ethernet switches 
are accessible through SNMP, HTTP, or command line interface.
It is possible to remotely configure VLANs and their associated spanning trees,
IGMP snooping and interface rate limits using management protocols.
It is also possible to remotely monitor switches for failures
and different activities can be triggered in reaction to these failure
events.  These features are thus referred to as programmable features of
modern Ethernet switches.

In this paper, we show that the programmable control mechanisms 
of modern Ethernet switches can be used to build 
the following high-level functionalities that are critical to 
MEN, SAN, and CI applications -- 
(i)~Enable traffic engineering that in turn routes
packet traffic to balance the load of physical network links. 
Routes obtained from switching path selection
are enforced by means of VLAN tags in a fashion similar to MPLS
labels. (ii)~Enable proactive switch and link disjoint backup path
provisioning to provide a very high degree of tolerance of  
switch or link failures. (iii)~Use rate limiting features of
Ethernet switches to regulate the bandwidth consumption of end nodes in
order to isolate different traffic from one another. (iv)~Use IGMP
snooping and link-layer multicast to support link-layer multicasting 
without broadcasting.

\figw{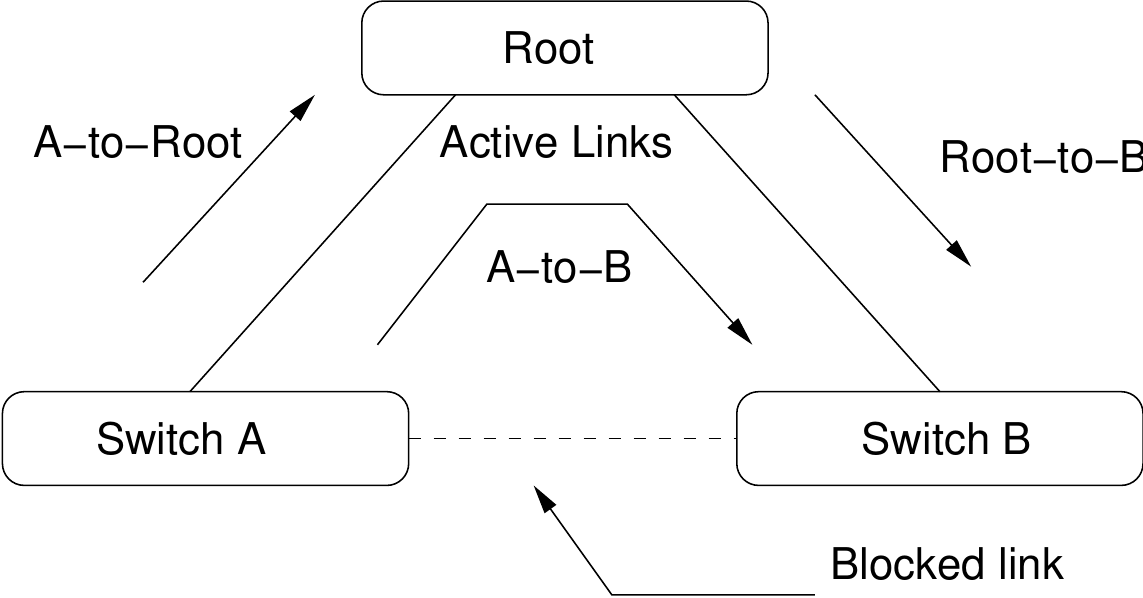}{2in}{Load imbalance scenario in spanning
tree based switching. Three different flows, A-to-B, A-to-Root, and
Root-to-B share same set of links despite the presence of a link between
switch A and B. If somehow flow A-to-B can be switched along link A-B, 
the overall network throughput can be improved significantly.}

\Se{Switching Path Selection}
In Ethernet, packets are always switched along the spanning tree of
the network.  The sender nodes\footnote{Throughout this paper, we use
the terms {\em nodes} and {\em hosts} to refer to the end-hosts that
are connected to a switched networks. The switch-nodes of the network
are always explicitly referred as {\em switches} in the paper.}  do not
have any control over the switching path.  In large scale networks like
MAN, the number of network elements (links and switches) involved in
a path between far off end-hosts is usually high.  Failure of any of
these elements can cause a complete communication breakdown between the
given pair of end-hosts.  Further, the intermediate links are shared by
multiple switching paths.  This sharing may lead to an overload situation
and the absence of alternate switching paths precludes any scope of load
balancing by offloading traffic to other links.  \mycomment{This would
happen even if redundant links are provided in the network because of
the very nature of STP to put the redundant links in blocking state.}
Figure~\ref{load-imbalance:fig} depicts a typical load imbalance scenario
when switching is done along a spanning tree.

Storage and cluster networks primarily  require  high
switching bandwidth and low communication latency between different
nodes. In switched Ethernet, the peak bandwidth between any two segments
is limited by the bandwidth of the link connecting them.  Addition of
multiple links does not improve the situation because, in spanning
tree topology, there can be only one active link forwarding traffic
to a segment. Thus, the general tendency is to aggregate the cluster
nodes on a single switch with high port density.  The aggregation cost
of switching networks increases rapidly with the size of the cluster
which in turn raises the per port cost factor.

\begin{figure}
\begin{center}
{\sf
{\small
\begin{tabbing}
abc\=foo\=foo\=foo\=foo\=foo\=foo\=foo\=foooooooooooo\=\kill
\>\> Let the set of all load balanced paths be $P$ \\
\>\> Let the set of all edge pairs be $EP$ \\
\>\> Let the set of spanning trees be $S$ \\
\>\> Let $S$ = $\phi$ \\
\>\> Sort the members of $P$ in the descending order of path \\
\>\> length \\
\>\> While ($EP$ $!=$ $\phi$ and $P$ $!=$ $\phi$ ) \\
\>\> Sort the members of $EP$ in descending order of their \\
\>\> frequency of  appearance in members of $P$ \\
\>\>\> Let $ep$ = Next element in $EP$ \\
\>\>\> While $\exists$ $p$ $\in$ $P$ such that $ep$ $\subset$ $p$ \\
\>\>\>\> Remove $p$ from $P$ \\
\>\>\>\> Find $s$ $\in$ $S$ such that $p$  and $s$ do not form a loop \\
\>\>\>\> Merge  $p$ with $s$ \\
\>\>\>\> If no such $s$ is found, add $p$ to $S$ \\
\end{tabbing}
}
}
\end{center}
\vspace{-0.1in}

\caption{\small Path Aggregation algorithm. For the given input of
selected paths $P$ and the network topology, the algorithm computes
a set of spanning trees. Each spanning tree can be associated with a
unique VLAN.}

\label{alg:aggregation}
\end{figure}

\Sse{VLAN-based switching} 

IEEE 802.1s MST protocol allows for the existence of multiple spanning
trees in an Ethernet network, where each spanning tree corresponds to
different VLANs. These multiple spanning trees can be used to provide load
balanced and fault-tolerant switching paths for different communicating
nodes.  Given the traffic profile or traffic requirements of any network,
one can come up with different load balanced switching paths such that
the overall network utilization is efficient~\cite{gopalan03efficient}.
These load balanced switching paths can be aggregated together such that
there are no loops in the aggregation. Such a constrained aggregation
would yield multiple spanning trees. Each spanning tree can further be
associated with a unique VLAN tag. The details of deriving load balanced
switching paths and grouping them into multiple spanning trees are
discussed in~\cite{viking}.  The path aggregation algorithm is shown in
Figure~\ref{alg:aggregation} for quick reference. This algorithm tries to
minimize the overall  number of spanning trees and hence VLANs required
after aggregation. Since the number of overall spanning trees required
is inversely proportional to the number of paths grouped together,
the number of spanning trees can be reduced by including larger number
of paths together. To increase the number of paths per spanning tree,
the algorithm tries to merge paths which share {\em common features}
such as sub paths, edges, or simply nodes. The end result would be
that each load balanced path is part of a spanning tree which is identified
by a unique VLAN tag.

For every incoming packet, the switches analyze the Ethernet header
for VLAN tags.  If a VLAN tag is found, the packet is switched along
the corresponding spanning tree.  Thus, any specific switching path can
be selected by simply inserting the appropriate VLAN tag in the packet
header. A host can insert different VLAN tags while communicating with
different destination nodes and thus can select different load balanced
switching paths.  This mechanism is very much analogous to MPLS, where
the packet switching paths are selected based on the labels present in
packet headers. The difference is that, in MPLS, the labels are inserted
by ingress routers, whereas here the VLAN tags need to be inserted by
the end hosts.

This VLAN-based switching is feasible only if a desired VLAN spanning
tree can be imposed (configured) on any network. This is where the
programmability features of Ethernet switches come into picture. Almost
all switches, which provide support for 802.1s MST protocol, also
facilitate programmability of links in terms of associated VLANs. Usually,
whenever VLANs are associated with different switches (links) in a
switched network, the switches participate in a distributed spanning tree
setup process and build a packet forwarding spanning tree. However, if the
links with which a particular VLAN is associated already form a spanning
tree, this spanning tree becomes the default switching spanning tree for
that VLAN. Thus, with the availability of VLAN programmability features in
switches, it is possible to realize this programmable VLAN-based switching
which enables efficient traffic engineering.

\figw{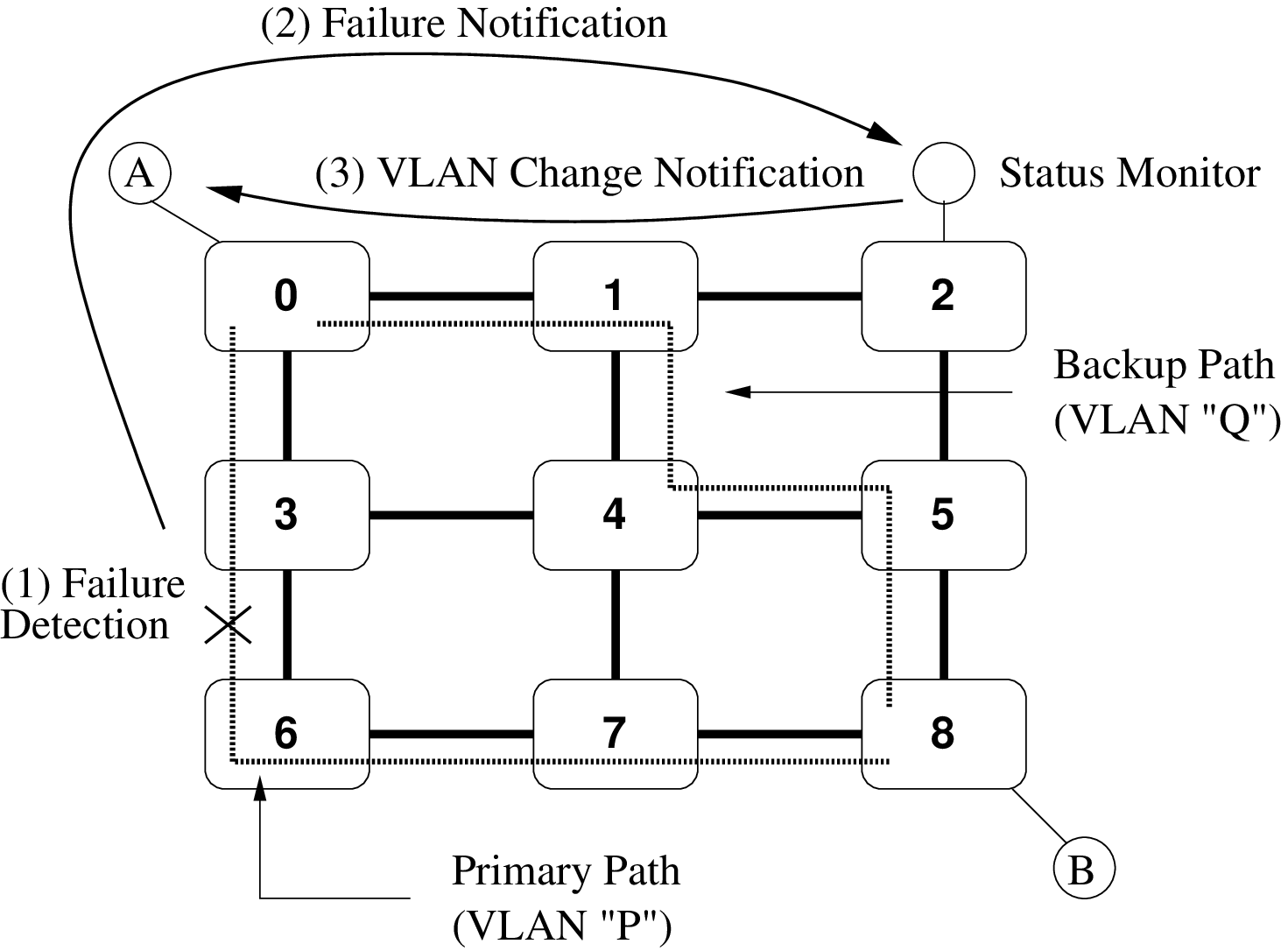}{2.2in}{Event monitoring based failure recovery mechanism.
The Status Monitor registers with all switches (0-9) for event
notifications. Node A and Node B communicate with each other using
path 0-3-6-7-9 over VLAN P. The failure of link 3-6 is detected by
switch 3 and the status monitor is notified about it. The status
monitor determines the list of all affected sender nodes and sends
them notifications to start using alternate VLANs.  In this case,
Node A receives a notification to use alternate VLAN Q which provides
the path 0-1-4-5-9. The entire recovery period comprises of failure
detection, failure notification to the status monitor, status monitor
lookup for alternate VLANs, and notification to the affected nodes.}

\Sse{Fault Tolerance}

Most managed (programmable) switches provide status monitoring facilities
where one can remotely setup traps which can get triggered  by various events like,
link failure, neighboring switch failure, link recovery, switch recovery,
neighbor discovery, etc. Using these traps one can detect topology changes
and failures in the network. If for every communicating host-pair, a backup
path is provisioned, it is easy to deal with failure scenario by simply
switching the communication over to the VLAN that corresponds to the
backup path.  In order to deal with failure of any switch or link in the
communication path, the backup path must be a link and switch disjoint
path to the primary path.  Due to the disjoint nature,  the primary and
backup paths belong to different spanning trees (and hence different
VLANs). Failures in one spanning tree do not affect other spanning trees
which do not include the failed links. Since the primary and backup paths
are pre-provisioned through traffic engineering, the fail-over duration
is limited to failure detection and event communication latencies between
switch and the status monitoring nodes.  The failure detection latencies
for commercially available managed switches such as Cisco Catalyst 2924
range from 400 to 500 milliseconds.  The fail-over period observed in
this case does not exceed 500 to 600 milliseconds and is consistently in
the sub-second range.  This is a significant improvement over multiple
second fail-over latency of 802.1w RSTP deployment. Note that, 802.1w
failure recovery period does not take into account the failure detection
period and is just the convergence period for spanning tree recovery.

One caveat about the above mentioned detection and recovery mechanism
is that it needs the presence of  status monitoring node(s) in the
network which is(are) aware of the entire network topology and the traffic
provisioning therein. Further, it is essential that the status monitoring
node is always reachable from all the switched in the network and all
nodes which need to be notified about the failure and the new VLAN tags
that need to be used.  If a network failure  disrupts the communication
path between the failure detecting switch and the monitoring node
itself, the fail-over mechanism cannot proceed.  However, this situation
can be overcome by dispatching the failure notification over multiple
communication paths to the status monitoring node. Another possibility
is to have multiple status monitoring nodes strategically located in
the network such that at least one status monitoring node can communicate
with  the failure detecting switch and all the affected nodes. An example
failure recovery scenario is depicted in Figure~\ref{backup:fig}.

\Se{Efficient Reliable Multicast}

Nodes in any high end cluster interconnect need to communicate with
each other for exchanging data as well as perform state checkpointing
and message logging for fault tolerance. It is often the case that
the same data is transmitted to multiple nodes during checkpointing and
message logging. For example, the cluster nodes may exchange periodic
heartbeat messages with  several other nodes monitoring the health of the
entire cluster.  The state checkpointing data may be replicated across
multiple nodes for an increased level of fault tolerance.  In storage
area networks, data is usually mirrored over multiple storage nodes. In
all these cases, the network performance can be significantly improved if
such communication is done over a multicast protocol rather than unicast
protocol. Traditionally, link-layer multicast over Ethernet networks was
implemented through coarse broadcast-and-filter semantics. The Ethernet
switches treated link-layer multicast packets as broadcast packets and the
transmission was done over all the switch port.  The onus of filtering out
irrelevant broadcast packets on the basis of destination MAC addresses
was placed on the end-host network interface cards (NICs). Though this
mechanism relieved end-host CPUs from redundant packet processing burden,
the network performance suffered significantly because of unnecessary
packet forwarding to all the ports.

Fortunately, most modern Ethernet switches, particularly Gigabit switches,
support a feature called {\em IGMP snooping}, which was designed to
support IP multicast without using link-layer broadcasting. In the IP
multicast schema, a node can join or leave a multicast group by sending
a {\tt join} or {\tt leave} IGMP~\cite{igmp} packets to the designated
routers which take care of propagating these messages to upstream routers.
In order to avoid redundantly broadcasting  the downstream multicast
traffic, Ethernet switches monitor these IGMP {\tt join} and {\tt leave}
messages to determine the ports on which the multicast traffic needs to
be forwarded. Subsequently, all downstream multicast traffic is forwarded to
only specific ports.

Although IGMP snooping was designed to improve the efficiency of IP
multicast on switched Ethernet networks, it can be exploited to support
link-layer multicast as well. Whenever a multicast group rooted at a
particular node needs to be created, the only thing the participating
nodes need to do is to send an IGMP {\tt join} message to the root
node. The intermediate switches would monitor the IGMP messages and
construct a spanning tree connecting the participating ports. Once such
a spanning tree is established, all multicast packets are forwarded along
the spanning tree rather than link-layer broadcasting.  If different such
spanning trees, catering to different multicast groups, are pre-configured
through traffic engineering for efficient network performance, all these
trees can be aggregated together in a way similar to path aggregation
described in Figure~\ref{alg:aggregation}.  This aggregation would yield
system-wide spanning trees.  These spanning trees can then be constructed
in the network by using the programmable features mentioned earlier.

The reliability of the link-layer multicast can be ensured by implementing
a positive acknowledgment and timeout based scheme. In this scheme,
after every multicast transmission, the sender sets a timer and expects
every receiver in the group to return an acknowledgment before
the timer expires. If some receivers fail to acknowledge the reception
in time, the sender can resort to unicast transmission to each such
receiver. This scheme is simple and efficient, and is optimized for the
case when packet drops and corruptions are extremely rare, as in the
case of switched Ethernet networks.


The number of link-layer multicast groups a switch can support is
limited.  Typically, the commodity switches can support up to
500 multicast groups.  To perform traffic engineering based on the
communication profile of all the nodes and to limit the number of
per-switch multicast groups  are some of the key challenges in multicast
provisioning. These problem can be addressed  by maximizing the number of
nodes that can share a multicast tree without affecting the overall performance
and then performing traffic engineering.  There can be various
ways the traffic engineering can be performed. One of the plausible
ways is to use the {\em hose} model~\cite{hose} developed for resource
management in VPNs. Though this model is challenging to support from
traffic provisioning aspect, it captures the essence of multicast traffic
in the sense that the traffic profile specification is in the form of
the amount traffic moving in and out of the nodes rather than pairwise
statistics as required by the traditional pipe models.


\Se{QoS Enforcement}
Traffic engineering and bandwidth provisioning alone cannot provide
quality of service in a network. It requires network usage policing
and regulation to provide a desired degree of QoS in the network.
Without usage regulation, the best one can do is to provide QoS
on the lines of priority based DiffServ~\cite{diffserv} in the
Internet. This kind of mechanism is inadequate to insulate different
traffic flows from one another and is not an acceptable scenario in
MAN where adherence to service level agreement (SLA) is a very crucial
requirement. Ethernet also specifies a DiffServ like traffic class
prioritization mechanism~\cite{8021p} which is supported by almost
all Ethernet switches. But it is clearly inadequate because of lack of
global enforcement.

The key to regulate the network usage and insulate different traffic flows
from one another is to monitor and enforce the usage right at the ingress
points in the network. If the amount of inflow network traffic does
not exceed the total engineered traffic, different traffic flows cannot
affect each other unless some of the network elements have failed. If
the traffic engineering is performed with proper backup provisioning, it
can be ensured that all the traffic flows are insulated from each other even
after failures.

Many mid-end and high-end Gigabit Ethernet switches support programmable
rate limiting features. For efficient wire speed,
these features are usually implemented in hardware. Rate limiting
can be used to provide restricted bandwidth usage based on predefined
profile or per physical port usage. Excess traffic can either be dropped
or reprioritized.  Though typically supported programmable parameters
of rate limiting are quite extensive, it can be simply specified  in
terms of the raw bandwidth limitations and the burst size limits. Using
these rate limiting features, it is possible to regulate the inflowing
traffic at the ingress ports. The amount of allowed inflow traffic
can be determined from the SLA in metro network scenario or from the
traffic characterization in the cluster interconnect scenario. Further,
the 802.1p traffic classification and prioritization can be used to mark
lower priority for the traffic that is sent in excess of the allowed
traffic so that  this  traffic is serviced only if there is spare
bandwidth available on every link and every switch along the path.

\Se{Programmability Requirements}
To realize the programmable switching Ethernet paradigm there are
certain minimum features that should be supported by the switches in
the network. The most important feature is the 802.1q tag based VLAN
support with 802.1s MST support. This feature needs to be supported
by all the switches in the network.  Similarly, all switches also
need to support link and switch status monitoring capability for fault
tolerance. Further, for a robust fault tolerance mechanism, the status
monitoring node needs to be able to communicate with all the switches
even after multiple failure.  This can be achieved by sending multiple
failure notification over different paths. Another alternative can be
to place multiple status monitors at strategic locations in the network.

The rate limiting feature is required to regulate the inflow of traffic
from end-hosts. It is worth exploring,  whether  all switches in the network need
to support this feature. Rate limiting is essential only when an end-host is
connected to a switch. In MAN setup, the switches can be classified
as core switches and edge switches where the edge switches serve as
ingress points. The core switches need not participate in the rate
limiting activity as the edge switches ensure that the traffic reaching
core switches is already rate limited. This mechanism is analogous to
ingress filtering in Internet.  An alternate viewpoint can be that,
usually there is spare bandwidth available at the edges but the core
of the network carries most of the traffic, so it should be the core
where bandwidth regulation takes place rather than at the edges. This
argument also has some merit in it.  However, the final decision can
be made only after traffic engineering. But unfortunately, traffic
engineering requires the knowledge of topology which usually gets
acquired after deployment. Moreover, when bandwidth provisioning is done
based on customer requirements and SLAs, service providers prefer to use
statistical multiplexing to maximize the network utilization.  The rate
limiting feature cannot deal with congestion occurring because of inherent
inability of statistical multiplexing to deal with worst case scenario.

In cluster and storage networks, potentially all switches may support
end-hosts. This makes it essential to support rate limiting at all
the switches. Also, the traffic profile may change from time to time.
Strict rate limiting on a changing traffic profile fails to capture
the changing requirement and hence is inappropriate. The proper way
to tackle this is to adapt soft rate filtering with re-prioritization
of excess traffic so that no traffic is dropped as long as there is
network capacity available to service it.  In addition, if the current
traffic profile differs significantly from the profile used for bandwidth
provisioning, traffic re-engineering needs to be carried out

Throughout the discussion, we implicitly assumed the availability of
traffic monitoring and engineering mechanism. Traditional pipe model
based traffic engineering requires pairwise load statistics for all
end-hosts. Even if traffic specifications is provided by customers, as in
MAN case, precise traffic profiling is required to exploit statistical
multiplexing. Commodity Ethernet switches maintain statistics such that
the amount of traffic entering and leaving the ports and some other
finer details such as traffic per VLAN, amount of multicast traffic, etc.
This is clearly insufficient to build a complete pairwise traffic profile.
There are two possible solutions to this problem.  The traffic engineering can use
the  {\em hose model}~\cite{hose} of resource management. This model 
provides the flexibility of traffic specification in terms of required 
input and output capacities at the end-hosts rather than a pairwise
traffic matrix. However, resource provisioning becomes a tougher problem
to solve in this context. An alternative is to modify the end-hosts to keep
track of pairwise traffic between all peers. These end-hosts can then
periodically communicate these statistics to the nodes responsible for
traffic engineering. This approach is feasible in storage and cluster
networks where all the nodes fall under a common administrative domain.

Whenever end-hosts need to communicate with other peers, the switching
path is selected by specifying the VLAN corresponding the spanning tree
along which the packets are switched. The sending node needs to insert the
appropriate VLAN tag in the packet header. This tag needs to be obtained
from some node aware of traffic engineering. Further, during fail-over,
the status monitors inform the sender nodes to change over to backup
VLANs.  This requires all senders to be aware of the path-selection
and fail-over mechanism in order to comply with and utilize them. It
may not be possible to modify the network stack of customer nodes in
environments like MAN.  In such cases, the service provider can address
this problem by placing proxy nodes which take care of interacting with
service provider nodes.




\mycomment{
\Sse{Application Specific Requirements}
\Ssse{Metro Ethernet}
\Ssse{Storage Area Networks}

\Ssse{Cluster Interconnects}
}

\Se{Performance Benefits}
We studied the improvement in aggregate through of a VLAN-based
multi-spanning-tree Ethernet architecture over its single spanning
tree counterpart using simulations. The simulations were carried out
to determine the maximum bandwidth that could be supported in the
network. The network topology was assumed to be a grid topology which
can represent metro Ethernets and cluster networks. The simulations were
carried out against a uniform traffic pattern of each node communicating
with other nodes with equal traffic load.
The uniform traffic distribution is a representation of cluster networks.
The simulations were run against grids of sizes 16, 25, 36,
49, and 64 nodes connected using links with a capacity of 100 Mbps. 
The traffic between nodes was 10, 8, 5, 2, and 1 Mbps for grids of sizes 16, 25, 36, 49, and 64 respectively.  
The effectiveness of path selection was compared for possible
throughput against the single spanning tree case. The path selection
was carried out for both cases, with and without backup redundancy. All
traffic comprised of at least one hop between the switches.

\figw{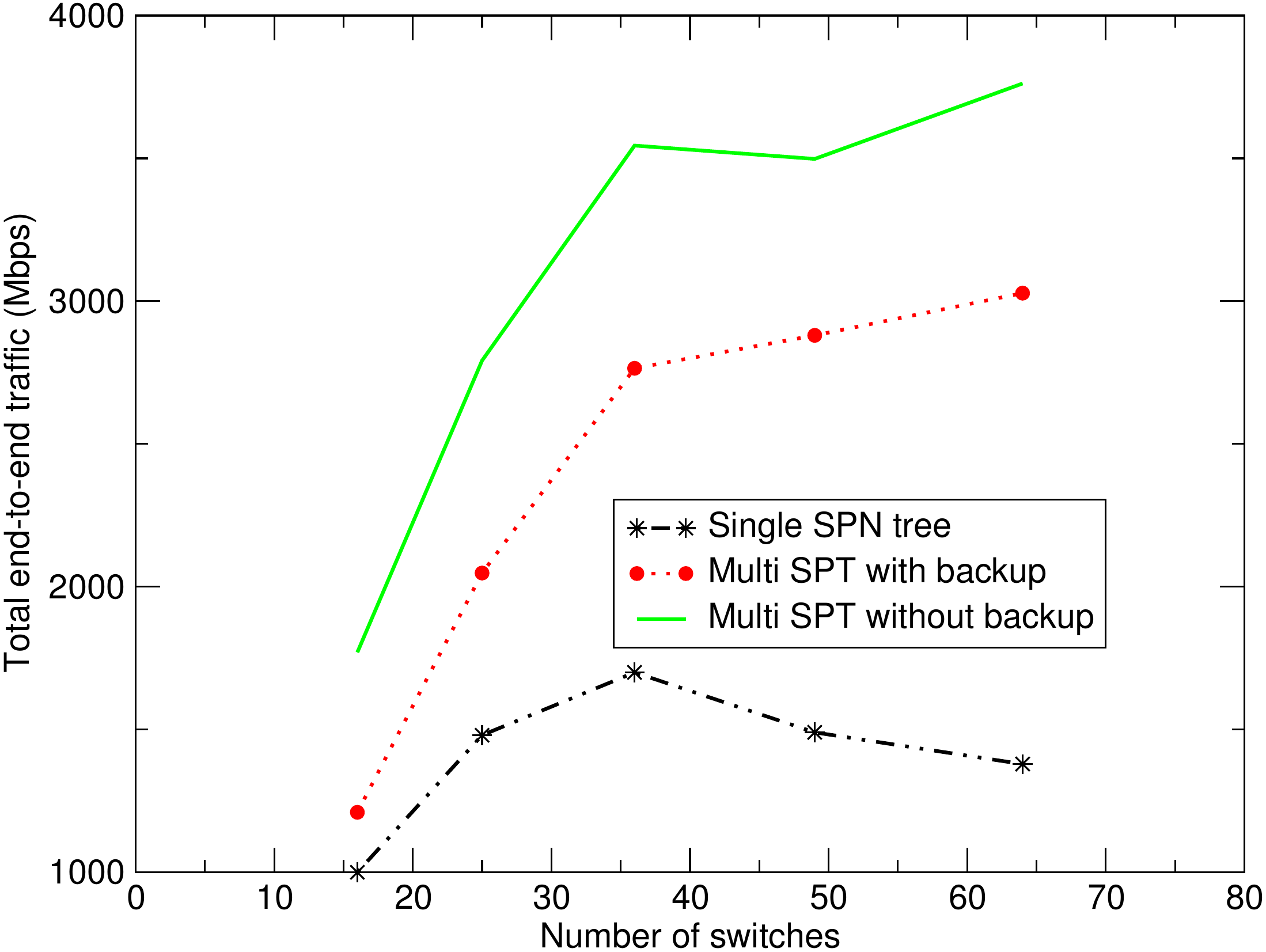}{2.2in}{ Total end-to-end traffic in network with
single spanning tree, and multiple spanning trees with and without
backup provisioning. The network follows a uniform traffic pattern while
communicating with each other.  }

Figure~\ref{traffic-all:fig} shows the comparative maximum throughput
for single spanning tree network against the VLAN-based switching path
selection. The traffic pattern is assumed to have a uniform distribution
across all node pairs. This is a representative distribution of cluster
networks. It can be seen that the total aggregate end-to-end throughput
is always more in system. As the number of nodes increases the performance
shows considerable increase. This is because of availability of additional
number of active links in topology.

\figw{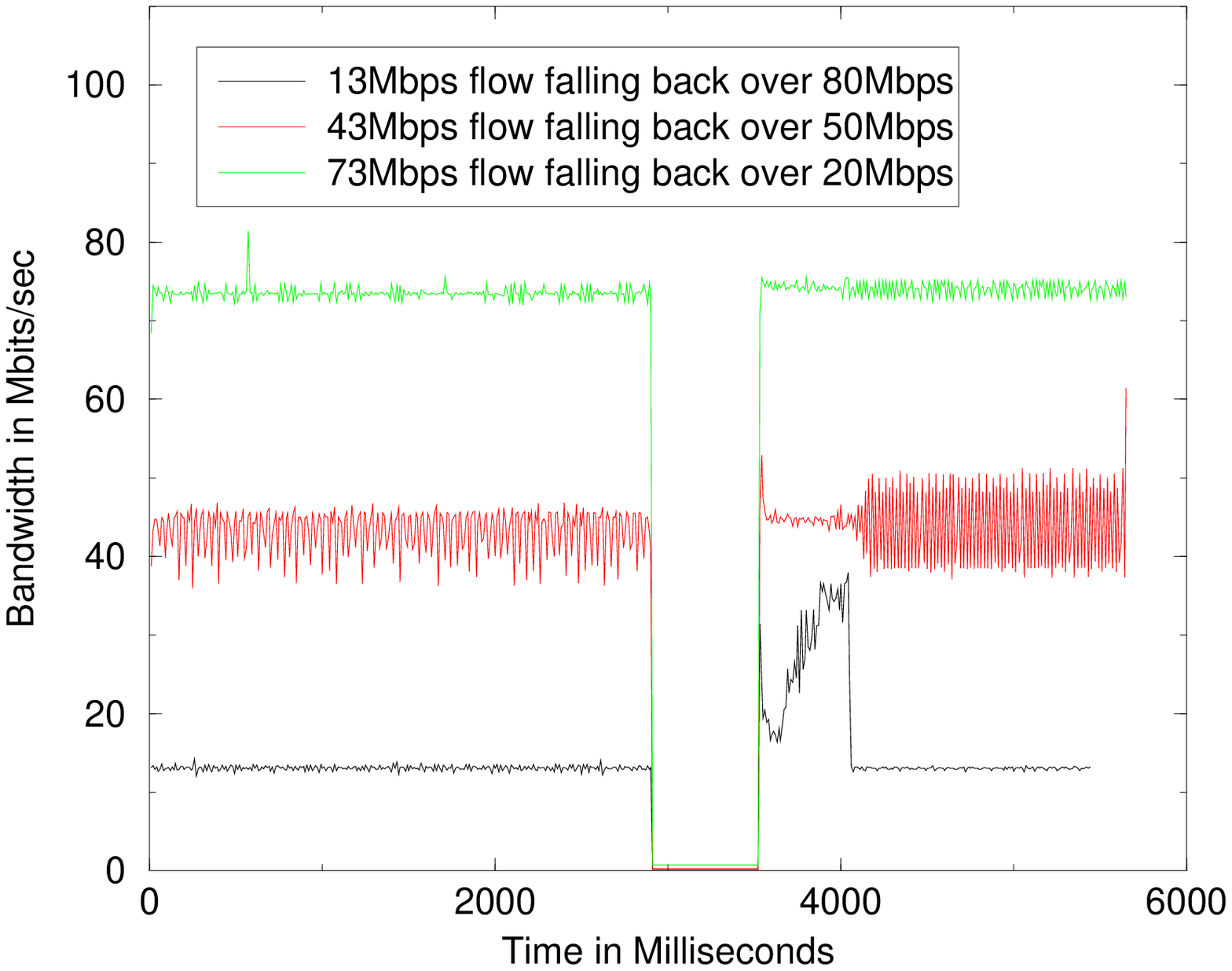}{2.2in}{Behavior of TCP across fail-over. After fail-over,
TCP has to adapt to the existing traffic on backup links. The amount of
traffic on backup links was maintained at a sufficiently high level so
that the links are saturated after fail-over.  This was the worst case
scenario for fail-over.  It was observed that TCP takes around 300 ms
to 400 ms to recover from fail-over.}

Figure~\ref{tcp:fig} shows the effect of link failure on TCP throughput.
The experiment was run with a setup where two switches were connected by
two links.  These links were then configured to belong to two different
VLANs. To have a more realistic scenario, we introduced enough background
traffic to keep the link utilization at the maximum level in the VLAN
that served the main traffic.  Upon link failures, the TCP traffic passing
through this link falls back onto the backup VLAN, which happens to pass
through the other link.  There was an expected drop of bandwidth for the
recovery period (around 600 milliseconds) Once the backup VLAN was used,
the TCP flow regained its momentum hardly suffering any more delay and
attaining stability in around 300ms to 400ms.
Thus, the {\em complete} failure recovery time can be broken down
into following components: The failure detection time, including the
event notification to status monitor, of around 400 to 600 milliseconds and
the VLAN change notification time of a few milliseconds.  The total down-time
incurred is in sub-second range.

%
%

\Se{Conclusion}
Lack of fine-grained path selection mechanism in Ethernet networks 
is a major barrier for the application of traffic engineering technology. 
The fallout of this restriction has been a limiting
factor for Ethernet deployment in the core carrier networks. Though Ethernet is becoming more and more
popular, this popularity is from access perspective and not from the core network
perspective. In this paper, we described how the programmable 
control mechanisms in modern Ethernet switches support  
MPLS-like load-balancing path selection 
to provide multi-fold network throughput gains. When combined with 
status monitoring in Ethernet, this fine-grained path selection can be used to provide
sub-second level fault tolerance, 
which is a required feature in core networks. 
We also described a simple efficient reliable multicast protocol which uses Layer-3 awareness
of commercially Ethernet 
switches to improve the performance of storage and cluster networks.
We believe there are many more high-level functionalities 
that can be composed from the basic programmable control mechanisms.
One of the future directions for this work is to explore and build
interesting high-level functionalities that can enhance large scale
networks as a whole.
\bibliography{switching}
\bibliographystyle{IEEE}

\end{document}